\documentclass[preprint,12pt]{elsarticle}

\usepackage{multirow,textcomp}
\usepackage{graphicx,booktabs}
\usepackage[small]{caption}
\usepackage{subfig}
\usepackage{amsmath,amsfonts,amssymb}
\usepackage{natbib}
\usepackage{bm}
\bibpunct{(}{)}{;}{a}{,}{,}

\journal{Journal of Statistical Planning and Inference}

\begin{document}

\begin{frontmatter}



\title{Branch and Bound Algorithms for Maximizing Expected Improvement Functions}


\author{Mark Franey, Pritam Ranjan and Hugh Chipman}
\ead{mark.franey@gmail.com, pritam.ranjan@acadiau.ca, hugh.chipman@acadiau.ca}
\address{Department of Mathematics and Statistics, \\ Acadia University, NS, Canada B4P 2R6}

\begin{abstract}
Deterministic computer simulations are often used as a replacement for complex physical experiments.  Although less expensive than physical experimentation, computer codes can still be time-consuming to run.  An effective strategy for exploring the response surface of the deterministic simulator is the use of an approximation to the computer code, such as a Gaussian process (GP) model, coupled with a sequential sampling strategy for choosing design points that can be used to build the GP model.  The ultimate goal of such studies is often the estimation of specific features of interest of the simulator output, such as the maximum, minimum, or a level set (contour). Before approximating such features with the GP model, sufficient runs of the computer simulator must be completed.

Sequential designs with an expected improvement (EI) function can yield good estimates of the features with a minimal number of runs.  The challenge is that the expected improvement function itself is often multimodal and difficult to maximize.  We develop branch and bound algorithms for efficiently maximizing the EI function in specific problems, including the simultaneous estimation of a minimum and a maximum, and in the estimation of a contour.  These branch and bound algorithms outperform other optimization strategies such as genetic algorithms, and over a number of sequential design steps can lead to dramatically superior accuracy in estimation of features of interest.
\end{abstract}


\end{frontmatter}


\renewcommand{\baselinestretch}{1.3}
\section{Introduction} \label{introduction}
Computer experiments are often employed to simulate complex systems,
but realistic computer simulators of physical phenomena can be time
consuming to evaluate. In such cases it is desirable to perform the
experiment as efficiently as possible. One popular approach
designed for this purpose is sequential experimental design
\citep{jonesreview} whereby an initial design is simulated and a
surrogate model is fit to the output, often using a Gaussian Spatial
Process \citep{gp1}. The surrogate provides estimates at all input combinations (sampled and unsampled) with accompanying variance estimates
and is used, along with a suitable `infill criterion', to choose
subsequent trials.

Scientists are often interested in specific features of the simulator, such as the maximum, the minimum, level set(s), etc.  In
each case, the follow-up trial in the sequential design is selected so as to
give the greatest improvement to the estimate(s) of these feature(s)
of interest.  A popular approach is to choose this new trial by
maximizing a merit-based infill criterion.  Many criteria have been
proposed \citep{jones} \citep{informationimprovement}
\citep{forrester2008global} and one popular choice is the `expected
improvement'. This criteria has been shown to be efficient if the
initial design is not too sparse or deceptive \citep{forrester2008global}.

The form of the improvement function varies with the features of
interest, but in any case facilitates a balance between choosing points that exhibit
local optimality (greedy search, new trials close to predicted
feature of interest), and points with high uncertainty (global
search, new trials away from previously sampled points)
\citep{infillcriteria}.  Finding the next design point (the point that maximizes the EI function) is necessary in order to ensure the
estimation of features of interest in the smallest number of
simulator runs.  Finding this maximum is a difficult global
optimization problem, and the addition of each point in the sequential design changes the EI function, requiring another search for the global optimum of EI in order to find the next point.

The main focus of this paper is to develop branch and bound
algorithms for maximizing expected improvement functions for two
particular features of interest; contours, and the simultaneous estimation of maximum and
minimum.  The most challenging aspect of this
algorithm is constructing the bounds on the objective function
(expected improvement). Branch and bound algorithms have been implemented for finding the
minimum of a function \citep{jones}, and expected improvement
criteria have been developed for the minimum \citep{jones}, and
contours \citep{ranjan}. Our main contribution are in (a) generalizing
the expected improvement criterion for simultaneous estimation of the maximum and
minimum, and (b) developing branch and bound
algorithms for maximizing expected improvement for contours, and for
simultaneous maximum/minimum estimation. \citet{ranjan} maximized their expected
improvement function for estimating contours using a genetic
algorithm. The branch and bound algorithm developed here
outperforms their optimization by locating level sets using fewer
simulator runs.

This paper is organized as follows: Section \ref{overview} gives an
overview of the Gaussian process model, the expected improvement
criterion and a generic branch and bound algorithm which will be used to
maximize the expected improvement over candidate input points.  In
Section \ref{main_main} we propose bounds on the expected improvement for
estimating contours and simultaneous estimation of the global maximum and minimum.
Section \ref{results} presents simulation results comparing our
branch and bound and a genetic algorithm similar to that of
\citet{ranjan} for different test functions. We present our conclusions and outline
possible future work in Section \ref{conclusion}.

\section{Review - surrogate model, global optimization} \label{overview}

\subsection{Gaussian Process Model}

Let $y_i$ represent the univariate response from the computer
simulator, and $\mathbf{X}=(\mathbf{x}_1, \mathbf{x}_2,...,
\mathbf{x}_n)'$ be the matrix of input vectors
$\mathbf{x}_i'=(x_{i1},x_{i2},...,x_{id})$ used to generate the
respective outputs, with  $y(\mathbf{X})=(y_1,y_2,...,y_n)'$.
Without loss of generality, let the input space be the unit
hypercube $\chi = [0,1]^d$. Following \citet{gp1} we model the
output $y(\mathbf{x}_i)$ as
\begin{equation}
 y(\mathbf{x}_i)=\mu + z(\mathbf{x}_i); \qquad i=1, ..., n,
\end{equation}
where $\mu$ is the overall mean, and $\{ z(\mathbf{x}), x \in
\chi\}$ is a Gaussian spatial process with mean 0, variance
$\sigma^2_z$ and corr$(z(\mathbf{x}_i),z(\mathbf{x}_j)) = R_{ij}$.
The correlation matrix $R$ is a function of the hyper-parameters $\bm{\theta} = (\theta_1, \ldots, \theta_d)$.
In general, $y(X)$ has multivariate normal distribution, $y(X) \sim
N_n({\bf 1_n}\mu, \sigma^2_zR)$, where the parameters $\Omega =
(\theta_1, ..., \theta_d; \mu, \sigma^2_z)$ are estimated by
maximizing the likelihood. The mean and variance parameters have
closed form estimates given by
\begin{equation}
\hat{\mu}=(\bm{1_n}' R^{-1}\bm{1_n})^{-1} \bm{1_n}' R^{-1}\bm{y} \quad \mathrm{and}   \quad   \hat{\sigma}^2_z=
\frac{(\bm{y}-\bm{1_n} \hat{\mu})'R^-1 (\bm{y}-\bm{1_n} \hat{\mu})}{n},
\end{equation}
whereas the hyper-parameters $\bm{\theta} = (\theta_1,...,\theta_d)$
are estimated by maximizing the profile likelihood.

The best linear unbiased predictor (BLUP) at $x^*$, $\hat{y}(x^*)$, and the associated mean squared error (MSE) $s^{2}(x^*)$ are given by
\begin{flalign}
\hat{y}(x^*) = \ &\hat{\mu}+r'R^{-1}(y-\mathbf{1}_n  \hat{\mu}), \\
s^2 (y(x^*)) = \ & \sigma_z^2\left( 1-r'R^{-1}r+
\frac{(1-\mathbf{1}'_n R^{-1}r)^2}{\mathbf{1}'_n R^{-1}\mathbf{1}_n}
\right),
\end{flalign}
where $r=(r_1(x^*),...,r_n(x^*))'$ and $r_i(x^*)=\mathrm{
corr}(z(x^*),z(x_i))$.  The estimates  $\hat{y}(x^*)$ and
$\hat{s}(x^*)$ are obtained by replacing $R$, $r$ and $\sigma^2_z$ with
$\hat{R}(\bm{\hat{\theta}})$, $\hat{r}(\bm{\hat{\theta}})$ and $\hat{\sigma}^2_z$.
There are innumerable choices for the correlation structure, for
instance, power exponential correlation and Mat\'ern correlation
(see \citet{stein1999, santner2003daa} for details).  The
choice of correlation function is not crucial to the algorithms and
methodologies developed here, and thus we omit the details. In all
illustrations, the power exponential correlation function is used.

It turns out that if the design points are close together in the input space the correlation matrix $R$ can sometimes be near-singular which results in unstable computation. A popular approach to overcome this problem is to introduce a small nugget $0< \delta < 1$ and
approximate the ill-conditioned $R^{-1}$ with a well-conditioned
$(R+\delta I)^{-1}$, where $I$ is the $n \times n$ identity matrix.
In later sections, a nugget is included in the GP model.


\subsection{Sequential design using EI criterion}
For expensive computer models the total number of
simulator runs is limited; and thus the choice of design points
is crucial for estimating certain pre-specified features or
approximating the underlying process. If we are
interested in only certain features (e.g., global maxima, contours,
and so on) of the process, a popular approach is to use a sequential
sampling strategy. The key steps of such a sampling strategy are
summarized as follows:
\begin{enumerate}
 \item Choose an initial design $\{x_1,...,x_{n_0}\}$.
 \item Run the simulator, $f$, for these design settings to get $\{y_1,...,y_{n_0}\}$, where $y_i = f(x_i)$. Set $n=n_0$.
 \item Fit a surrogate to the data $\{(x_i,y_i), i=1,...,n\}$ (we use the Gaussian process model).
 \item Choose a new trial $x_{new}$ that leads to improvement in the estimate of the feature of interest.
 \item Run the simulator for the new trial location and update the data $x_{n+1} = x_{new}$, $y_{n+1} = f(x_{new})$ and $n=n+1$.
 \item Repeat 3-5 until $n$ exceeds a predetermined limit.
\end{enumerate}

Developing feature specific criteria for selecting new trials has
gained considerable attention in the computer experiment community. For
instance,  \citet{jones} proposed an \emph{expected improvement}
(EI) criterion for estimating the global minimum of an expensive
computer simulator. Let $I(x)$ denote the improvement function for estimating a particular feature of interest. Then, the proposed criterion in \citet{jones} is $E[I_1(x)]$, where $I_1(x) = \max\{
f_{min}-y(x),0 \}$, $f_{min}$ is the current best estimate of the
function minimum, and the expectation is taken over the distribution
of $y(x) \sim N(\hat{y}(x),s^2(x))$. That is, the new trial in Step~4 is the
maximizer of
\begin{equation}\label{ei_min}
E[I_1(x)] = s(x)\phi(u) + (f_{min}-\hat{y}(x))\Phi(u),
\end{equation}
where $u = (f_{min}-\hat{y}(x))/s(x)$, and $\phi(\cdot)$ and
$\Phi(\cdot)$ denote standard normal pdf and cdf respectively. One of the most attractive characteristics of the EI criterion is that it exhibits a balance between `global' and `local' search. The first term in $(\ref{ei_min})$ supports global search, and the second term, local search.

Since the main objective of performing a good design is to attain
(or at least reach a good approximation of) the global minimum in as
few simulator runs as possible, efficient optimization of the EI
function becomes very important. The EI functions are often multimodal
and the location, number and heights of these peaks change after adding
every new trial. \citet{jones} developed a branch and bound (BNB)
algorithm for optimizing the EI function (\ref{ei_min}), which we
review briefly in the next section.

\subsection{Branch and bound algorithm: review for finding a global minimum} \label{optimization}
BNB algorithms are often used for global optimization of non-convex
functions \citep{lawler1966branch,moore1991global}. The BNB algorithm
presented here finds the global minimum of a real-valued function
$g$. In our case, $g(x) = -E[I(x)]$ is defined on the
$d$-dimensional hypercube $\mathcal{Q}_{init} =\chi= [0,1]^d$ for a specific feature of interest.

There are three key components of a BNB algorithm - (i) branching
(ii) bounding and (iii) pruning. The branching component uses a splitting
procedure that, given a rectangle $\mathcal{Q} \subset
\mathcal{Q}_{init}$, returns $\mathcal{Q}_I$ and $\mathcal{Q}_{II}$
such that $\mathcal{Q}_I \cap \mathcal{Q}_{II} = \phi$ (null) and
$\mathcal{Q}_I \cup \mathcal{Q}_{II} = \mathcal{Q}$. The splitting
occurs along the longest edge of $\mathcal{Q}$ (if there is a tie,
one is chosen randomly). Bounding is the second component.  Bounding the objective function $g$, requires finding the lower and upper bounds, $\Psi_{lb}$ and $\Psi_{ub}$, of the minimum value of $g$. Let $\Psi_{min}(\mathcal{Q}) = \min_{x \in \mathcal{Q}} g(x)$ for every $\mathcal{Q} \subset \mathcal{Q}_{init}$, then $\Psi_{lb}$ and $\Psi_{ub}$ must satisfy
\begin{eqnarray}\label{r1_r2} \nonumber
R1 \quad &:&  \Psi_{lb}(\mathcal{Q}) \leq \Psi_{min} (\mathcal{Q}) \leq \Psi_{ub}(\mathcal{Q}) \\
R2 \quad &:&  \forall \; \epsilon > 0 \; \exists \; \delta > 0 \; \textrm{such that for all} \; \mathcal{Q} \subseteq \mathcal{Q}_{init}, \\ \nonumber
 & & | \mathcal{Q}| \leq \delta \; \Longrightarrow \; \Psi_{ub}(\mathcal{Q})-\Psi_{lb}(\mathcal{Q}) \leq \epsilon.
\end{eqnarray}
As in \citet{bnb}, $| \mathcal{Q}|$ is the length of the longest edge of rectangle $\mathcal{Q}$. Although BNB algorithms guarantee the global minimum of $g$ with a
pre-specified tolerance $\epsilon$, the bounding functions $\Psi_{lb}$ and $\Psi_{ub}$ are objective function specific and often nontrivial to derive. This is the
most challenging part of a BNB algorithm. Appropriate lower and upper bounds can be constructed for maximizing the EI criteria for a few specific process features of
interest, and this is the main contribution of the paper. The third component of the BNB algorithm is pruning,
in which the algorithm removes rectangles $\mathcal{Q}$ from the set of all
rectangles $\mathfrak{L}$, that have lower bound greater than the
upper bound of some other rectangle in $\mathfrak{L}$. In the BNB algorithm outlined below, individual rectangles are represented by $\mathcal{Q}$'s and lists of rectangles are represented by $\mathfrak{L}$'s:\\

\noindent
1 $k=0$ \hspace*{1in} (initialize counter) \\
2 $\mathfrak{L}_{0}=  \{\mathcal{Q}_{init}\}$ \hspace*{1in} (initialize rectangle list)\\
3 $L_{0} =  \Psi_{lb}\{\mathcal{Q}_{init}\}$ \\
4 $U_{0} =  \Psi_{ub}\{\mathcal{Q}_{init}\}$ \\
5 while $U_k-L_k> \epsilon$\\
6 \hspace*{15 pt}$Pick \; \mathcal{Q} \in \mathfrak{L}_k \; : \; \Psi_{lb}(\mathcal{Q})=L_k$\\
7 \hspace*{15 pt}$Split \; \mathcal{Q} \; along \; one \; of \; its \; longest \; edges \; into \; \mathcal{Q}_I \; and \mathcal{Q}_{II}$ \\
8 \hspace*{15 pt}$ \mathfrak{L}_{k+1} := (\mathfrak{L}_k \backslash \{\mathcal{Q}\}) \cup \{\mathcal{Q}_I,\mathcal{Q}_{II}\} $  \hspace*{0.5in} (remove $\mathcal{Q}$, replace with \\ \hspace*{3.5in} $\mathcal{Q}_I$ and $\mathcal{Q}_{II}$ to get $\mathfrak{L}_{k+1}$)\\
9 \hspace*{15 pt}$ L_{k+1} := min_{\mathcal{Q}\in \mathfrak{L}_{k+1}} \Psi_{lb}(\mathcal{Q})$  \\
10 \hspace*{10 pt}$U_{k+1} := min_{\mathcal{Q}\in\mathfrak{L}_{k+1}} \Psi_{ub}(\mathcal{Q})$  \\
11 \hspace*{10 pt}$ \mathfrak{L}_{k+1} := \mathfrak{L}_{k+1} \setminus \{\mathcal{Q} \in \mathfrak{L}_{k+1}:\Psi_{lb}(\mathcal{Q})>U_{k+1}\}$\\
12 \hspace*{10 pt}$k=k+1$ \\
13 $end$\\

\noindent In this algorithm, $k$ is the iteration index, $\mathfrak{L}_{k+1}$ is the
list of hyper-rectangles, $(L_{k+1},U_{k+1})$ are the smallest lower and upper bounds
for $\Psi_{min}(\mathcal{Q}_{init})$ (in our case $\min_{x \in \chi}- E[I(x)]$)
after $k+1$ iterations, and $\epsilon$ is the desired precision and is
fixed beforehand. Steps~6-8 correspond to branching, Steps~9-10 bounding, and Step~11 represents pruning of the hyper-rectangles.

The EI function, $E[I_1(x)]$, for finding the global minimum of the
simulator, is a function of the input location $x$ via the predicted response $\hat{y}(x)$,
the associated uncertainty $s^2(x)$ at $x$ and the current
estimate of the global minimum $f_{min}$ (in general the estimate of the feature of
interest). \citet{jones} note that $E[I_1(x)]$ is monotonic with respect to (w.r.t.) $\hat{y}(x)$ and $s(x)$. Let $EI_1(s(x), \hat{y}(x))$ denote $E[I_1(x)]$ for all $x \in \chi$. Taking partial derivatives of $(\ref{ei_min})$ w.r.t. $\hat{y}(x)$ and $s(x)$ give
\begin{equation*}\label{ei_min_partials}
\frac{\partial EI_1}{\partial s(x)} = \phi \left( u \right)
\quad \textrm{and} \quad \frac{\partial EI_1}{\partial
\hat{y}(x)} =  -\Phi \left(u \right).
\end{equation*}
The partial derivatives $\partial EI_1/\partial s(x) \geq 0$ and
$\partial EI_1/\partial \hat{y}(x) \leq 0$ for all $x \in \chi$.
Hence if $\hat{y}(x)$ and $s(x)$ can be bounded by
$(\hat{y}_{lb}(\mathcal{Q}), \hat{y}_{ub}(\mathcal{Q}))$ and
$(s_{lb}(\mathcal{Q}), s_{ub}(\mathcal{Q}))$ respectively over a
hyper-rectangle $\mathcal{Q} \in \mathfrak{L}$, then for every $x
\in \mathcal{Q}$, the lower and upper bounds $E[I_1(x)]_{ub} =
\Psi_{lb}(\mathcal{Q})$ and $E[I_1(x)]_{lb} =
\Psi_{ub}(\mathcal{Q})$ are
\begin{flalign*}
E[I_1(x)]_{lb} & = EI_1(s_{lb}(\mathcal{Q}),\hat{y}_{ub}(\mathcal{Q})),\\
E[I_1(x)]_{ub} & = EI_1(s_{ub}(\mathcal{Q}),\hat{y}_{lb}(\mathcal{Q})),
\end{flalign*}
where $\hat{y}_{lb}(\mathcal{Q}), \hat{y}_{ub}(\mathcal{Q}),
s_{lb}(\mathcal{Q})$ and $s_{ub}(\mathcal{Q})$ are the lower and
upper bounds on $\hat{y}(x)$ and $s(x)$ over  $\mathcal{Q}$. These
bounds are needed in Steps~6-11. In practice $s(x)$ is replaced by
its predicted value $\hat{s}(x)$. This approach of bounding the EI
function is efficient, as bounding $s(x)$ and $\hat{y(x)}$ is
relatively easier (see \citet{jones} for details).


\section{BNB Algorithms for New EI Criteria}\label{main_main}

In this section we first propose a generalization of the expected
improvement criterion developed by \citet{jones} for simultaneously
estimating the maximum and minimum of an expensive deterministic
computer simulator. Next we develop a BNB algorithm for maximizing
this new EI criterion. We also propose a modification in the EI
criterion developed by \citet{ranjan} for estimating a pre-specified
contour. This modification facilitates the development of a BNB
algorithm for maximizing the modified EI criterion, and still
maintains the global versus local search trade-off.


\subsection{EI criterion for maximum and minimum} \label{main}
We propose an improvement function for simultaneous estimation of
the global maximum and minimum. This feature could be of specific
interest, for instance, if one wishes to identify best-case and
worst-case scenarios in a global climate change model, or maximal and minimal projected
unemployment rates in a complex model of the economy. The
improvement function can be written as
\begin{equation}
I_2(x)=\max\{(y(\mathbf{x})-f_{max}),(f_{min}-y(\mathbf{x})),0\},
\end{equation}
where $f_{max}$ and $f_{min}$ are current best estimates of the
global maximum and minimum respectively. The corresponding expected
improvement criterion is obtained by taking the expectation of $I_2(x)$
with respect to the distribution of $y(x) \sim N(\hat{y}(x),
s^2(x))$,
\begin{equation}\label{ei_maxmin}
E[I_2(x)] = s\phi(u_1) + (\hat{y}-f_{max})\Phi(u_1) + s\phi(u_2) +
(f_{min}-\hat{y})\Phi(u_2),
\end{equation}
where $u_1 = (\hat{y} - f_{max})/s$ and $u_2 = (f_{min} -
\hat{y})/s$. The EI criterion (\ref{ei_maxmin}) turns
out to be the sum of the two EI criteria for individually estimating
the global maximum and the global minimum. Optimization of (\ref{ei_maxmin}) favors subsequent sampling in the
neighbourhood of the global maximum if $(\hat{y}-f_{max})\Phi(u_1)$ is the dominating
term in the sum, the global minimum if $(f_{min}-\hat{y})\Phi(u_2)$ is the dominant
term, and otherwise in the less explored regions to minimize overall variability.


As in optimization of the EI criterion (5) for estimating the
global minimum, a BNB algorithm can be developed for optimizing (\ref{ei_maxmin}). The most difficult part, computation
of the lower and upper bounds of (\ref{ei_maxmin}) that satisfy $R1$ and $R2$ in (\ref{r1_r2}) for minimizing $g(x) = -E[I_2(x)]$, is accomplished by monotonicity of the $E[I_2(x)]$ w.r.t. $\hat{y}(x)$ and $s(x)$. Let $EI_2(s(x), \hat{y}(x))$ denote $E[I_2(x)]$ for all $x \in \chi$. Then the two partial derivatives of $E[I_2(x)]$ are
\begin{flalign*}
\frac{\partial  EI_2}{\partial s(\mathbf{x})} = \phi \left( u_1
\right)+\phi \left( u_2 \right) \quad \textrm{and} \quad
\frac{\partial EI_2}{\partial \hat{y}(\mathbf{x})} =  \Phi
\left( u_1 \right) -\Phi \left(u_2 \right).
\end{flalign*}
The partial derivative w.r.t. $s(x)$ is positive for all $x \in
\chi$, and the partial derivative w.r.t. $\hat{y}(x)$ is equal to
zero when $\hat{y}(\mathbf{x}) = \frac{(f_{max}+f_{min})}{2}$.
Moreover, it is straightforward to show that
\begin{displaymath}
\frac{\partial EI_2}{\partial \hat{y}(\mathbf{x})} \left\{
\begin{array}{ll}
 \ge 0 &  \textrm{if}\ \hat{y}(\mathbf{x}) > \frac{(f_{max}+f_{min})}{2} \\
 \le 0 &  \textrm{if}\ \hat{y}(\mathbf{x}) < \frac{(f_{max}+f_{min})}{2}\\
 \end{array} . \right.
\end{displaymath}
Hence $E[I_2(x)]$ is monotonic for all $x \in \chi$ w.r.t. $s(x)$
and piecewise monotonic w.r.t. $\hat{y}(\mathbf{x})$. Given the
upper and lower bounds on $\hat{y}(\mathbf{x})$ and
$s(\mathbf{x})$ in a hyper-rectangle $\mathcal{Q} \subset \mathcal{Q}_{init}$, somewhat \emph{conservative} bounds on
$E[I_2(\bm{x})]$, for every $x \in \mathcal{Q}$, are
\begin{flalign}
E[I_2(x)]_{lb} = \min \{&EI_2(s_{lb}(\mathcal{Q}),\hat{y}_{lb}(\mathcal{Q})),
EI_2(s_{lb}(\mathcal{Q}),\hat{y}_{ub}(\mathcal{Q}))\}, \nonumber \\
E[I_2(x)]_{ub} = \max \{&EI_2(s_{ub}(\mathcal{Q}),\hat{y}_{lb}(\mathcal{Q})),
EI_2(s_{ub}(\mathcal{Q}),\hat{y}_{ub}(\mathcal{Q}))\} \nonumber .
\end{flalign}
That is, the upper bound of $E[I_2(x)]$ in $\mathcal{Q}$ is
calculated at $s_{ub}(\mathcal{Q})$, and due to piecewise
monotonicity of $E[I_2(x)]$, corresponds to the
$\hat{y}(\mathbf{x})$ in $\mathcal{Q}$ that is closest to $f_{min}$
or $f_{max}$. Similarly, the lower bound of $E[I_2(x)]$ is
calculated using $s_{lb}(\mathcal{Q})$ and the $\hat{y}(\mathbf{x})$
in $\mathcal{Q}$ that is closest to $(f_{max}+f_{min})/2$.

These bounds are conservative because the lower and upper bounds of
$s(x)$ and minimizer of $|\hat{y}(x) - (f_{max}+f_{min})/2|$ may not
correspond to the same point in the input space. The lower bound
$\Psi_{lb}(\mathcal{Q})$ of $\min_{x \in \mathcal{Q}} (-E[I_2(x)])$
will be equal to the true minimum of $-E[I_2(x)]$ only if the
minimizers of $s(x)$ and $|\hat{y}(x) - (f_{max}+f_{min})/2|$ are
identical. As a result, the hyper-rectangles in $\mathfrak{L}$ are
pruned at a slower rate.


\subsection{BNB algorithm for contour estimation problem} \label{EIcontour}
The use of a EI criterion for estimating a contour (level set) of an expensive computer simulator was proposed by \citet{ranjan}. The motivating application involved distinguishing ``good" from ``bad" performance of a server in a one-server-two-queue network queuing simulator. The proposed improvement function for estimating the contour $S(a) = \{x
: y(x) = a\}$ was $I_3(x) = \epsilon^2(x) - \min \{ (y(x)-a)^2,
\epsilon^2(x) \}$, where $\epsilon (x) = \alpha s(x)$, for a
positive constant $\alpha$, defines a neighbourhood around the contour. The expectation of $I(x)$ w.r.t. $y(x) \sim N(\hat{y}(x), s^2(x))$ is given by
\begin{flalign} \label{eicontour}
E[I_3(x)] = &\ \left[ \epsilon^2(x) -(a-\hat{y}(x))^2\right] \left[ \Phi
\left(u_1\right)-\Phi \left(u_2 \right) \right] \\
& -2(a-\hat{y}(x))s(x) \left[ \phi \left(u_1\right)-\phi
\left(u_2 \right) \right]
- s^2(x) \int_{u_1}^{u_2} w^2 \phi \left(w \right) dw, \nonumber
\end{flalign}
where $u_1 = (a - \hat{y}(x) + \epsilon(x))/s(x)$, $u_2 = (a - \hat{y}(x) - \epsilon(x))/s(x)$, and $\phi(\cdot)$ and $\Phi(\cdot)$ are the standard normal pdf and cdf respectively. \citet{ranjan} use $E[I_3(x)]$ as the EI criterion for selecting additional new trials in the sequential sampling strategy. However, controlling the trade-off between
local and global search has gained attention in the area of
computer experiments (e.g.,  Schonlau, Welch and Jones (1998),
S\'obester, Leary, and Keane (2005)). This motivated us to modify
the EI criterion in (\ref{eicontour}) that will facilitate the
construction of the lower and upper bounds which satisfy $R1$ and
$R2$ in (\ref{r1_r2}), and still entertains the global versus local search trade-off.


Note that the third term in (\ref{eicontour}) represents the total
uncertainty in the $\epsilon$-neighbourhood of the contour at $x$,
and can be dropped without altering the important features of the
criterion. We propose a modified EI criterion, obtained by dropping
the third term in (9) and using a change of variable $\{\hat{y}(x),
s(x)\} \rightarrow \{t(x), s(x)\}$, given by
\begin{flalign} \label{eicontour_new}
E[I_3^*(x)] = &\ s^2(x)\left[ \alpha^2 -t^2\right] \left[ \Phi
\left(t+\alpha\right)-\Phi \left(t-\alpha \right) \right]
\\\nonumber & -2ts^2(x) \left[ \phi
\left(t+\alpha\right)-\phi \left(t-\alpha \right) \right],
\end{flalign}
where $t = (a - \hat{y}(x))/s(x)$. Although the proposed
modification slightly alters the tradeoff between the global and
local search (see Section \ref{eiNewOldComparison} for more
details), it allows the partial derivatives of $E[I_3^*(x)]$ to be
piecewise monotone. Furthermore, this facilitates easy construction
of lower and upper bounds in the BNB algorithm for efficient
optimization of $E[I_3^*(x)]$.

\begin{figure} \centering
\includegraphics[width=0.8 \textwidth]{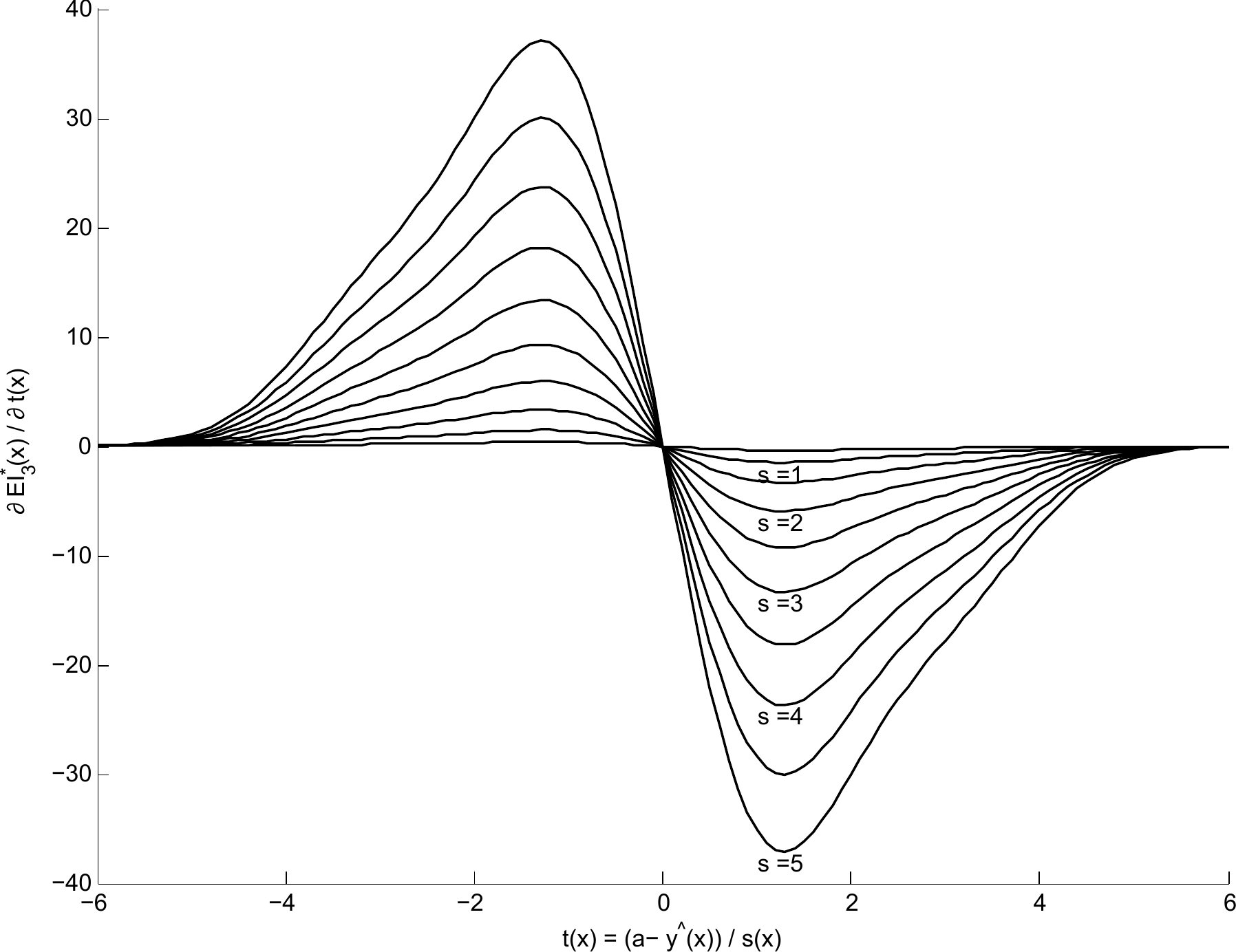}
\caption{Partial derivative of $EI_3^*(t(x), s(x))$ w.r.t. $t(x)$ evaluated at $\alpha=2$.}
\label{fig:partial_derivative_yhat}
\end{figure}

Let $EI_3^*(s(x), t(x))$ denote $E[I_3^*(x)]$ for all $x \in \chi$. Then, the partial derivative of $E[I_3^*(x)]$ w.r.t. $t=t(x)$ is
\begin{eqnarray}\label{eq:partial_t} \nonumber
\frac{\partial EI_3^*}{\partial t} &=&
-2s^2(x)t[\Phi(t+\alpha) - \Phi(t-\alpha)] + 2\alpha s^2(x)
t[\phi(t+\alpha) + \phi(t-\alpha)]\\ & & +\ s^2(x)(\alpha^2
+t^2-2)[\phi(t+\alpha) - \phi(t-\alpha)].
\end{eqnarray}
It turns out that the sign of the partial derivative $\partial EI_3^*/\partial t$ depends on the sign of $t=(a-\hat{y}(x))/s(x)$. Since the normal pdf is symmetric around the mean (i.e., $\phi(u) = \phi(-u)$, for all $u$), the partial derivative $\partial EI_3^*/\partial t = 0$ at $t=0$. In general,
\begin{displaymath}
\frac{\partial EI_3^*}{\partial t} \left\{
\begin{array}{ll}
 \le 0 &  \textrm{if}\ t > 0 \\
  =  0 &  \textrm{if}\ t = 0 \\
 \ge 0 &  \textrm{if}\ t < 0\\
 \end{array} . \right.
\end{displaymath}
Due to the presence of normal pdf and cdf in (\ref{eq:partial_t}), it may not be straightforward to analytically prove the monotonicity result, however, it can easily be plotted (see Figure~\ref{fig:partial_derivative_yhat}).

There are a few points worth noting. For instance, the height of these peaks, in Figure~1, increases with $s$. The magnitude and sign of the derivative depend on the choice of $\alpha$ (we used $\alpha =2$). Interestingly, the nature of the curves remain the same for $\alpha > 1$ (approximately).

\begin{figure} \centering
\includegraphics[width=0.8 \textwidth]{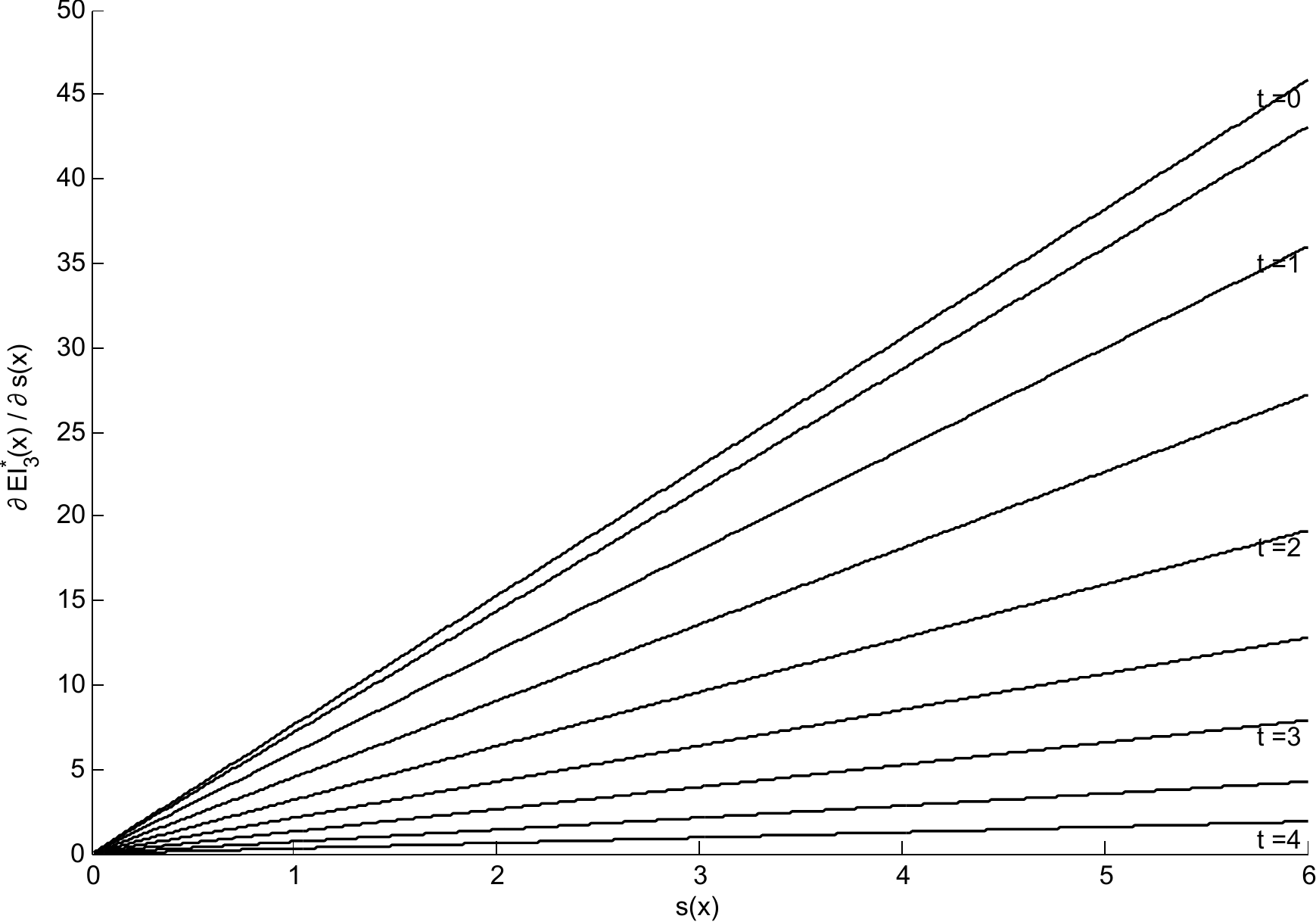}
\caption{Partial derivative of $EI_3^*(t(x), s(x))$ w.r.t. $s(x)$ evaluated at $\alpha=2$.}
\label{fig:partial_derivative_shat}
\end{figure}

The partial derivative $EI_3^*$ w.r.t. $s(x)$ is given by
\begin{eqnarray}\label{eq:partial_shat}\nonumber
\frac{\partial EI_3^*}{\partial s(x)} &=&
2s(x)(\alpha^2-t^2)[\Phi(t+\alpha) - \Phi(t-\alpha)] \\
& & - 4ts(x)[\phi(t+\alpha) - \phi(t-\alpha)].
\end{eqnarray}
The partial derivative $\partial EI_3^*/\partial s(x)$ is always non-negative. However,  as with the monotonicity of $EI_3^*(s(x), t(x))$ with respect to $t(x)$,
analytic proof may be somewhat complicated. Figure~\ref{fig:partial_derivative_shat} displays the partial derivative $\partial EI_3^*/\partial s(x)$ as a function of $s(x)$.

As in the simultaneous estimation of the global maximum and minimum case, bounds on $E[I_3^*(x)]$ for every $x \in \mathcal{Q}$ are obtained using the bounds on $t(x)$ and $s(x)$ in the hyper-rectangle $\mathcal{Q} \in \mathfrak{L}$. The bounds are
\begin{flalign*}
&E[I_3^*(x)]_{lb}  = \min \{EI_3^*(s_{lb}(\mathcal{Q}),t_{lb}(\mathcal{Q})),
EI_3^*(s_{lb}(\mathcal{Q}),t_{ub}(\mathcal{Q}))\}, \\
&E[I_3^*(x)]_{ub}  = \max \{EI_3^*(s_{ub}(\mathcal{Q}),t_{lb}(\mathcal{Q})),
EI_3^*(s_{ub}(\mathcal{Q}),t_{ub}(\mathcal{Q}))\}.
\label{eiubcontour}
\end{flalign*}
Because $EI_3^*(s(x), t(x))$ is an increasing function of $s(x)$, the lower bound of $E[I_3^*(x)]$ in $\mathcal{Q}$ is obtained using $s_{lb}(\mathcal{Q})$, and the value of $\hat{y}(x)$ farthest from the estimated contour.

\citet{ranjan} used a genetic algorithm for maximizing the expected improvement criterion $(\ref{eicontour})$. We will show in Sections~4.2 and 4.3 that the proposed BNB algorithm outperforms the implementation of \citet{ranjan} and can save simulator evaluations.

\section{Simulations} \label{results}

This section presents comparison between the proposed branch
and bound algorithm, a genetic algorithm (GA) and a non-sequential (static)
design for maximizing the EI criteria for various test functions. Two approaches are taken to compare BNB, GA and a static design approach. The first we will call the ``direct
comparison'', and was designed to reduce all noise external to the two
optimization algorithms (BNB and GA) so that the EI criterion values chosen by each
method can be directly compared. The second, the ``long
run comparison'', demonstrates the average performance of the two algorithms by
comparing the estimated and true features of interest as the
additional new trials are augmented. In both cases, complex test functions
are used for generating computer simulator outputs.

The GA we implemented to maximize the expected improvement function can be described as follows (see \citet{ranjan} for details):

\label{section:ga}
{\small
\noindent
\begin{verbatim}
for i=1..number_multistarts
   Generate an initial population X_init (of size n_init)
   for j=1..number_generations
     Mutation - Randomly perturb X_init to get X_pert
                Augmentation: X_aug = [X_init; X_pert]
     Crossover - Randomly swap elements of X_aug to get X_cross
                Augmentation: X_aug = [X_aug; X_cross]
     Selection - Evaluate EI for every candidate solution in
         X_aug and retain n_init members with highest EI value.
  end
end
\end{verbatim}
}
\noindent GAs are well-known algorithms that can produce competitive results. In the GA we used, the magnitude of the mutation was variable and perturbed each dimension $\pm$(0..5)$\%$ of the original value.  The initial population was generated with the random Latin hypercube function \texttt{lhsdesign} in Matlab 7.5.0. Crossover was between randomly selected pairs, in random dimensions.  See \citet{holland1975ana} and \citet{ga} for more information.

Before we compare the optimization power of the proposed BNB algorithms and the GA, we present results from an empirical study on the trade-off between local versus global search in $EI_3$ and $EI_3^*$, for estimating a pre-specified contour.

\subsection{Comparison between $EI_3$ and $EI_3^*$} \label{eiNewOldComparison}
Both $EI_3$ and $EI_3^*$ consist of terms that encourage local and
global search. The most important difference is the balance between
the local and the global searches. The entries in
Table~\ref{tab:EIs_contour_branin_levy} denote the proportion of additional trials that were chosen in the neighborhood of a pre-specified contour, i.e., the proportion of new trials that favored local search. Table~\ref{tab:EIs_contour_branin_levy} presents the results when the computer simulator outputs were generated from the Branin and two-dimensional Levy functions (see Section~4.3 for details on these test functions). The results presented here are averaged over 100 realization of random maximin Latin hypercube designs of size $n_0 = 20$ for initial design. For each realization, the new trials were chosen by maximizing the two EI criteria using GA. The contours of interest were $y=45$ and $y=70$ for the Branin and Levy functions respectively.  This simulation will display whether there are practical differences between the two criteria in terms of their balance between local and global search.

\begin{table}[hbt]\centering
\caption{Proportion of additional trials chosen in the neighbourhood of a pre-specified contour, by $EI_3$ and $EI_3^*$ criteria for estimating pre-specified contours}
\begin{tabular}{|c|cc|cc|}
\hline
number of & \multicolumn{2}{|c|}{Branin $y=45$ Contour} & \multicolumn{2}{|c|}{Levy 2D $y=70$ Contour}\\
points added & $EI_3$ & $EI_3^*$ & $EI_3$ & $EI_3^*$\\
\hline
$k=5 $ & 0.55 & 0.66 &0.49 &0.55 \\
$k=10$ &0.72 &0.80 & 0.44& 0.53 \\
$k=20$ & 0.85& 0.89&0.39 &0.54\\
$k=30$ & 0.89& 0.93& 0.37& 0.55 \\
\hline
\end{tabular}\label{tab:EIs_contour_branin_levy}
\end{table}

A new point was considered to be in the local neighbourhood if $f(x_{new})$ was within $(40,50)$ for the Branin function, and within $(60,80)$ for the Levy function.  For  both functions more points are added in the local neighbourhood with the new criterion,  most notably in estimating the Levy contour when $k=30$ new points were added.

The practical implications of this simulation are that the new
criterion allocates slightly more points for local search. This is
expected as the term with integral was removed from
(\ref{eicontour}), and it contributes to the global search.  This is
a desirable characteristic of an expected improvement criterion if
the underlying surface is relatively smooth.  If it is known that
the simulator response surface is relatively bumpy, the local and
global search terms in (\ref{eicontour_new}) could be reweighted as
in \citet{ei_weighted}.  This reweighted EI will still enable easy
construction of the lower and upper bounds for the branch and bound
algorithm.

\subsection{Direct comparison}
The objective of this simulation is to compare BNB and GA optimization of the EI criterion for a single step of the sequential design algorithm, starting from the same initial design.

To begin with we follow Steps~1-3 of the sequential design strategy outlined in Section~2.2 to obtain a fitted surface. The fitted GP model serves as input for the BNB and GA optimizer.
These two methods search for the maximum of the EI criterion over $\mathcal{Q}_{init}=[0,1]^d$, and then the estimates of the maximum EI are compared. The two optimizers BNB and GA use exactly the same inputs (MLE of the parameters in GP model, initial trials $X_{init}$, $y(X_{init})$, and the current best estimate of the feature of interest)
in this comparison, and so their output are directly comparable, though subject to small approximation. The approximation in BNB comes from estimating the bounds of $\hat{y}(\mathbf{x})$, $\hat{s}(\mathbf{x})$ based on points in each rectangle
$\mathcal{Q} \in \mathcal{Q}_{init} = [0,1]^d$. For example the upper bound of $\hat{y}(\mathbf{x})$ in some rectangle $\mathcal{Q}$ is estimated by the maximum observed $\hat{y}(\mathbf{x})$ for $\mathbf{x} \in \mathcal{Q}$ from the GP model fit. Thus, the comparison will be between a `stochastic version' of the branch and bound algorithm
developed in Section~3, and the genetic algorithm outlined earlier.
For all the comparisons presented here the number of evaluations of the EI criterion will be fixed for both BNB and GA; 500 for 2-dimensional, and 3000 for 4-dimensional test functions.  In actual applications, the use of a a tolerance $\epsilon$ for BNB might be preferred to the evaluation budget adopted here.  The evaluation budget facilitates direct comparison.

The entries in Tables~\ref{tab:bnb_ga_maxmin_contour} and 3 summarize the EI values found by BNB and GA for two features of interest (contour, maximum and minimum) and three test functions (Branin, two-dimensional and four-dimensional Levy). The numbers are the maximum EI values averaged over approximately 500 different initial random maximin Latin hypercube designs. A few simulations resulted in bad GP fit (especially when fitting the complex Levy functions, see Figure~6) and both BNB and GA results for those simulations were excluded from average calculation. The numbers in parentheses show the standard errors of these EI values ($s/\sqrt{n}=s/\sqrt{500}$). For example, in the first row of Table~2, 9.85 and 7.05 are the average maximum EI values obtained by BNB and GA, and the second row contains the corresponding sample standard error of 0.18 and 0.20 respectively.

\begin{table}[!ht]\centering \label{tab:bnb_ga_maxmin}
\caption{EI optimization using BNB and GA for simultaneous estimation of the global maximum and minimum, and contours for the Branin and 2D Levy functions}
\begin{tabular}{|c|cc|cc|}
\hline
\multicolumn{5}{|c|}{Maximum $\&$ minimum} \\
\hline
&\multicolumn{2}{|c|}{Branin}& \multicolumn{2}{|c|}{Levy 2D}\\
& BNB & GA &  BNB & GA  \\
\hline
$n_0 = 10$ & 9.85 & 7.05                &  3.58  & 2.74  \\
           & (0.18) & (0.20)    & (0.23) & (0.18)\\
$n_0 = 20$ &  7.46 &  4.45          &  1.33  & 1.13  \\
           &  (0.17) &  (0.18)  & (0.11) & (0.10)\\
$n_0 = 30$ &  6.51 &  3.48          &  0.76  &  0.66 \\
           &  (0.17) &  (0.16)  & (0.06) & (0.05)\\
$n_0 = 40$ &  5.61 &  2.75          &  0.68  &  0.66 \\
           &  (0.18) &  (0.14)  & (0.05) &  (0.04)\\
\hline
 \multicolumn{5}{|c|}{Contour at $y=a$}\\
\hline
&\multicolumn{2}{|c|}{Branin ($a=45$)}& \multicolumn{2}{|c|}{Levy 2D ($a=70$)}\\
& BNB & GA &  BNB & GA  \\
\hline
$n_0 = 10$ & 29.79 & 18.12                  & 62.70 & 41.88  \\
           & (0.67) & (0.49)      &(6.19) & (4.81)  \\
$n_0 = 20$ &  8.02 &  4.04                  & 55.64 & 40.23  \\
           &  (0.24) &  (0.15)      & (5.14) & (4.27) \\
$n_0 = 30$ &  3.54 &  1.90                  & 37.98 & 24.15 \\
           &  (0.15) &  (0.09)      & (3.58) & (2.57)  \\
$n_0 = 40$ &  1.76  &  0.93                 &  29.13 &  15.65  \\
           &  (0.08) &  (0.05)      &  (3.19) &  (1.99) \\
\hline
\end{tabular}\label{tab:bnb_ga_maxmin_contour}
\end{table}

Table \ref{tab:bnb_ga_maxmin_contour} shows that for every case
(choice of $n_0$, feature of interest and test function) BNB
maximizes $E[I_2(x)]$ more effectively than GA. In some cases the
two methods are closer - for instance with 40 initial points in
simultaneous estimation of the maximum and minimum for the 2D Levy
function the two methods are practically indistinguishable. But is
most of these simulations BNB locates significantly higher EI than
GA.
\begin{table}[!ht]\centering \label{tab:bnb_ga_4d_contour}
\caption{EI optimization using BNB and GA for estimation of the $y=180$ contour, and simultaneous estimation of the maximum and minimum for 4D Levy function}
\begin{tabular}{|c|cc|cc|}
\hline
 \multicolumn{5}{|c|}{4 Dimensional Levy Function }\\
\hline
&\multicolumn{2}{|c|}{$\qquad$Contour y=180$\qquad$ }& \multicolumn{2}{|c|}{Maximum and Minimum}\\
& BNB & GA  & BNB & GA  \\
\hline
$n_0 = 30$ & 516.1 & 509.8              & 11.79 & 10.63  \\
           & (38.39) & (38.53)    & (0.54) & (0.47)  \\
$n_0 = 40$ &  325.5 &  315.4        & 12.46 &  10.80      \\
           &  (27.14) &  (25.27)  & (0.56) &  (0.50)  \\
$n_0 = 50$ &  230.5 &  227.2        & 8.99 &  8.13    \\
           &  (16.62) &  (15.52)  & (0.38) &  (0.33)   \\
$n_0 = 60$ &  172.1 &  170.4        & 9.12 &  8.35    \\
           &  (14.90) &  (13.61)  & (0.39) &  (0.35) \\
\hline
\end{tabular}\label{tab:bnb_ga_4d_maxmin}
\end{table}


Analogous to the results presented in
Table~\ref{tab:bnb_ga_maxmin_contour},
Table~\ref{tab:bnb_ga_4d_maxmin} presents the results for the
four-dimensional Levy function. The features of interest are the
contour at $y=180$ and simultaneous determination of the maximum and
minimum. Results are presented when BNB and GA are allocated 3000
evaluations of the EI criterion
(Tables~\ref{tab:bnb_ga_maxmin_contour} and
\ref{tab:bnb_ga_4d_maxmin} respectively). For the 4D Levy function,
BNB achieves slightly better results than GA for both of these EI
criteria.

It should be noted that one of the principal advantages of the branch and bound algorithm is that it is designed to give results
within $\epsilon$ of the true maximum of $E[I(x)]$. Fixing the number of EI evaluations (and consequently the number of iterations of the BNB/GA optimization algorithm) though necessary for comparison, results in an approximation of the $EI(x)$ maximum which is not necessarily within $\epsilon$ tolerance of the true $E[I(x)]$ maximum.

\subsection{Long run comparison}
The maximum $EI$ values found by $BNB$ and $GA$ cannot be compared after the first new point is added because the augmented designs $\{X_{init} \cup x_{ga}^{new}\}$ and $\{X_{init} \cup x_{bnb}^{new} \}$ differ, as will the fitted GP surfaces and the EI surfaces. As a result, instead of comparing the maximum $EI$, the running best estimate of the features of interest (e.g., maximum) after adding $k$-th new design point are compared. We also use a non-sequential (static) approach, by fitting a GP model with a random maximin Latin hypercube design of size $n_0+k$, to serve as a benchmark for comparison. Ideally, both BNB and GA should outperform this non-sequential approach. Next, we present three examples to illustrate the comparisons.

\begin{figure}[!h]
  \centering
  \subfloat[Minimum]{\label{braninmaxminMin}\includegraphics[height=3in,width=2.5in]{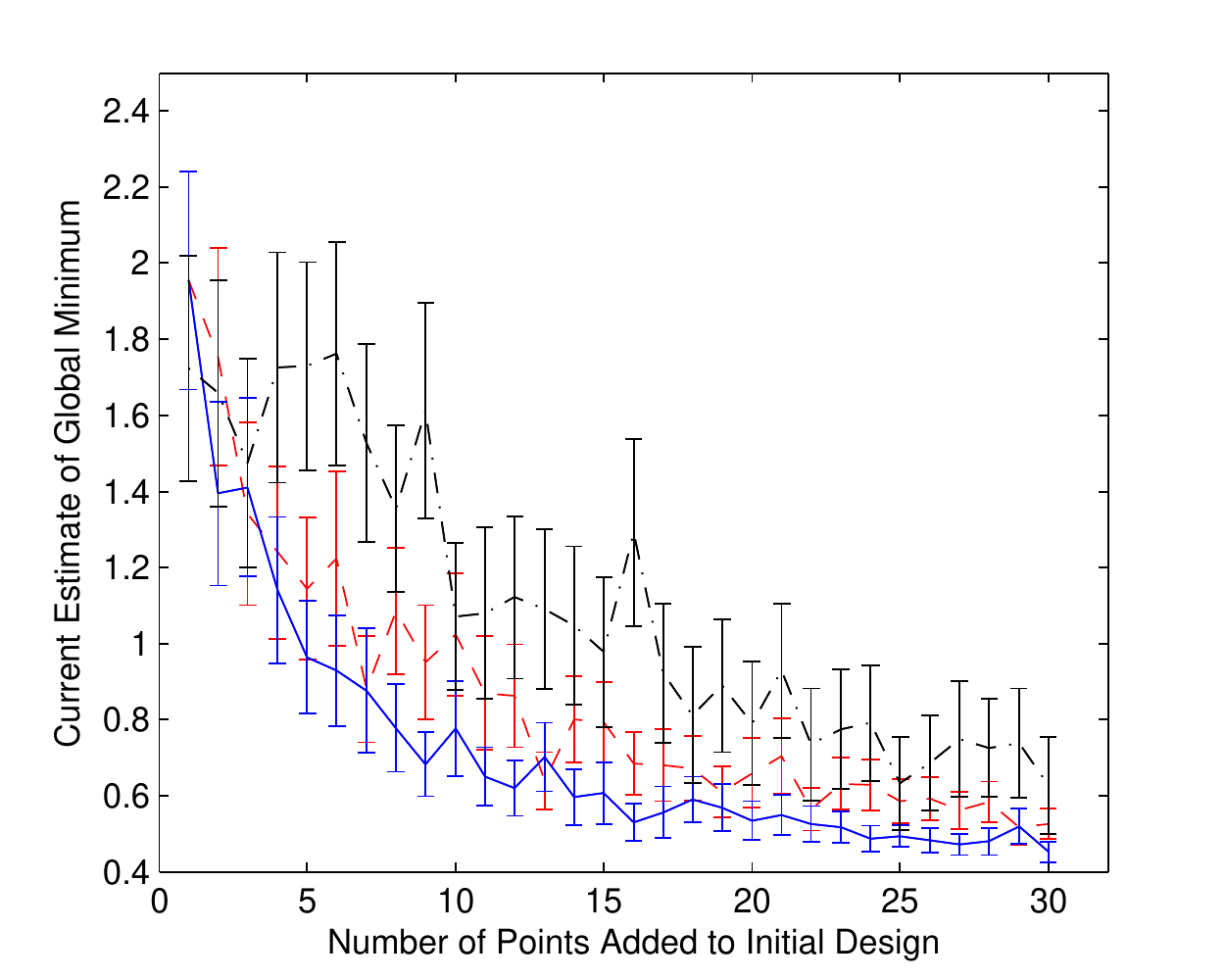}}
  \subfloat[Maximum]{\label{braninmaxminMax}\includegraphics[height=3in,width=2.5in]{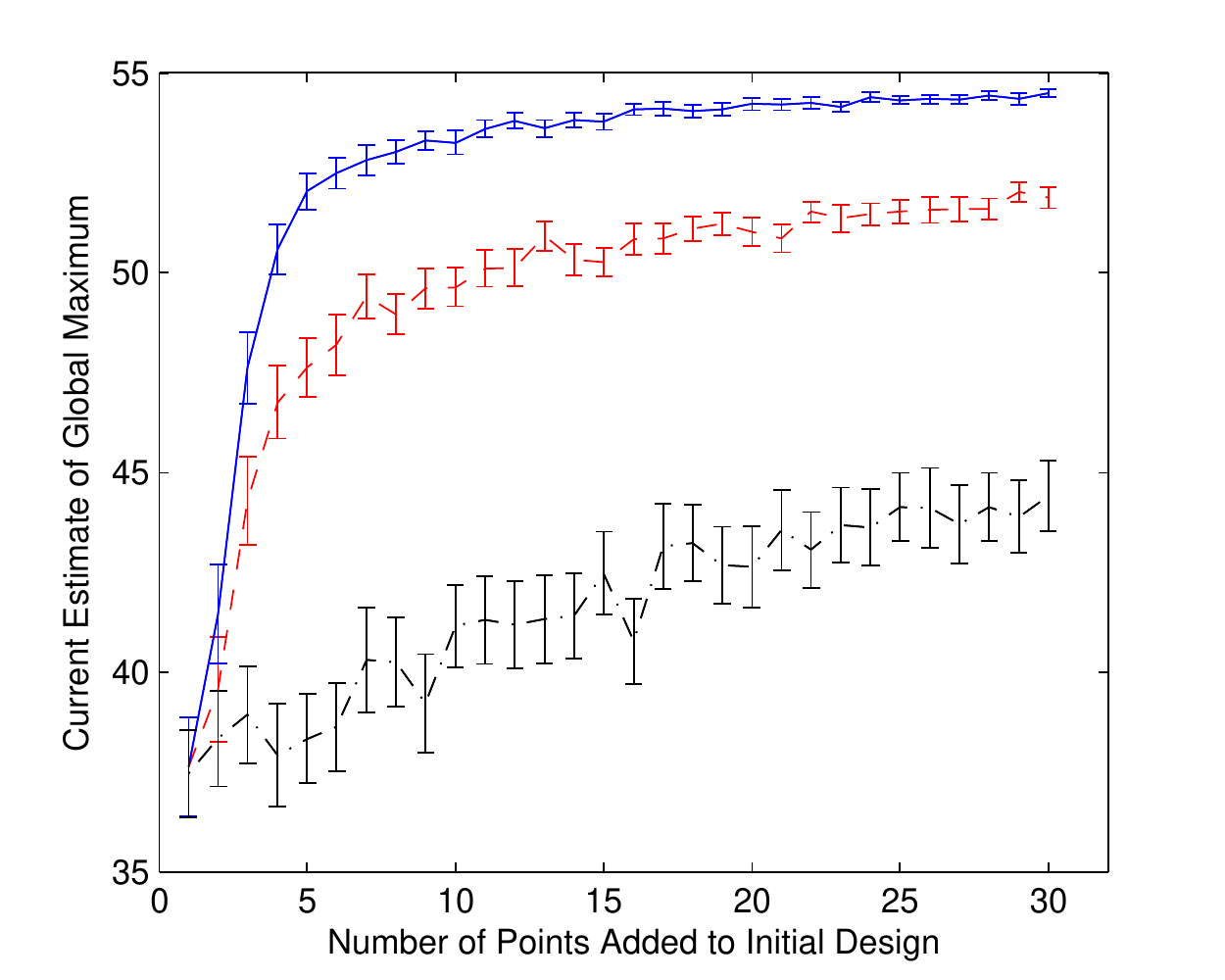}}
  \caption{Estimated optima for Branin function (true maximum = $55.6$, minimum = $0.398$);  BNB (blue - solid), GA (red - dashed), static (black - dash-dotted)}
  \label{braninmaxmin}
\end{figure}

\textbf{Example~1 - Branin.} Suppose the computer model output $y = f(x_1,x_2)$ is generated using Branin function given by
\begin{equation}
f(x_1,x_2) = \left(x_2-\frac{5.1 x_1^2}{4\pi^2} +\frac{5 x_1}{\pi}-6\right)^2+10 \left(1-\frac{1}{8 \pi}\right) \cos(x_1)+10,
\label{braninfunction}
\end{equation}
where $x_1, x_2 \in [0,5]$, however, we rescale the inputs to $[0,1]^2$. The true global maximum, $y_{max} =55.6$, and minimum, $y_{min} = 0.398$, of Branin function are attained at $(x_1,x_2)=(0,0)$ and $(x_1,x_2)=(0.62,0.42)$. Figure \ref{braninmaxmin} presents the long run comparison of the BNB and GA optimizers for maximizing the EI criterion (\ref{ei_maxmin}) developed for simultaneous estimation of the global maximum and minimum. For each realization, a random maximin Latin hypercube designs of size $n_0=20$ was chosen as initial design, and then $n_{new}=30$ additional trials were found sequentially (one at-a-time)
by maximizing the EI criterion and augmented to the design. The results are averaged over $100$ such realizations, and the error bars denote the simulation standard error.

\begin{figure}[!h] \centering
\includegraphics[height=3in,width=4.5in]{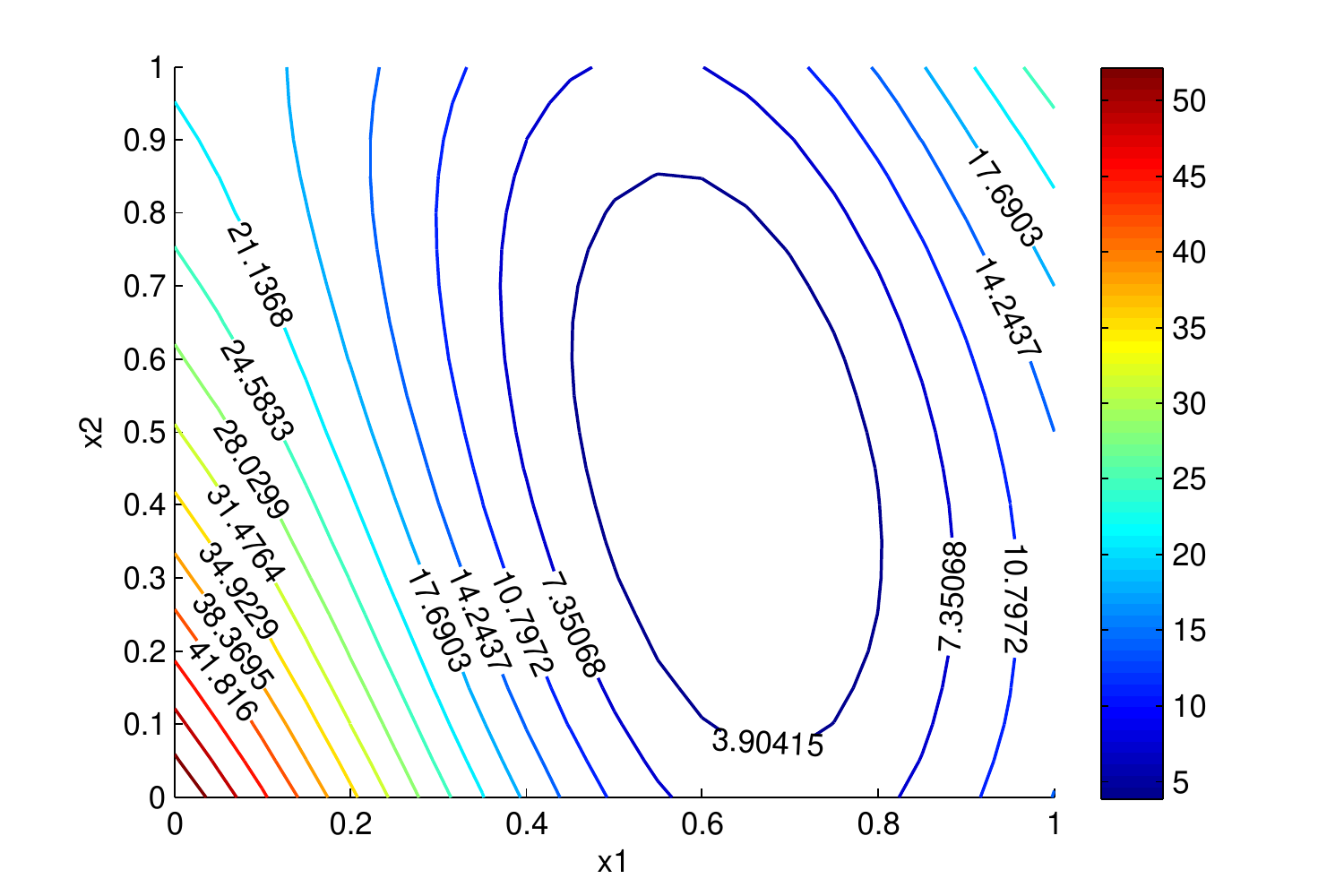}
\caption{Contours of Branin function in $[0,5]^2$ rescaled to $[0,1]^2$.}
\label{braninplot}
\end{figure}

It is clear from Figure~\ref{braninmaxmin} that the optimization of the EI criterion using BNB (blue) leads to the true maximum in significantly fewer number of computer simulator runs compared to the (red) GA optimizer or baseline (black) search of a random maximin Latin hypercube. However the estimated global minimum does not seem to be significantly different for any of the optimization techniques, though on average BNB attains a closer estimate of the true minimum than GA which is in turn closer on average than the static method. This is somewhat intuitive as the Branin function is quite smooth and almost flat near the global minimum (see Figure~\ref{braninplot}).

\begin{figure}[!h] \centering
\includegraphics[scale=0.66]{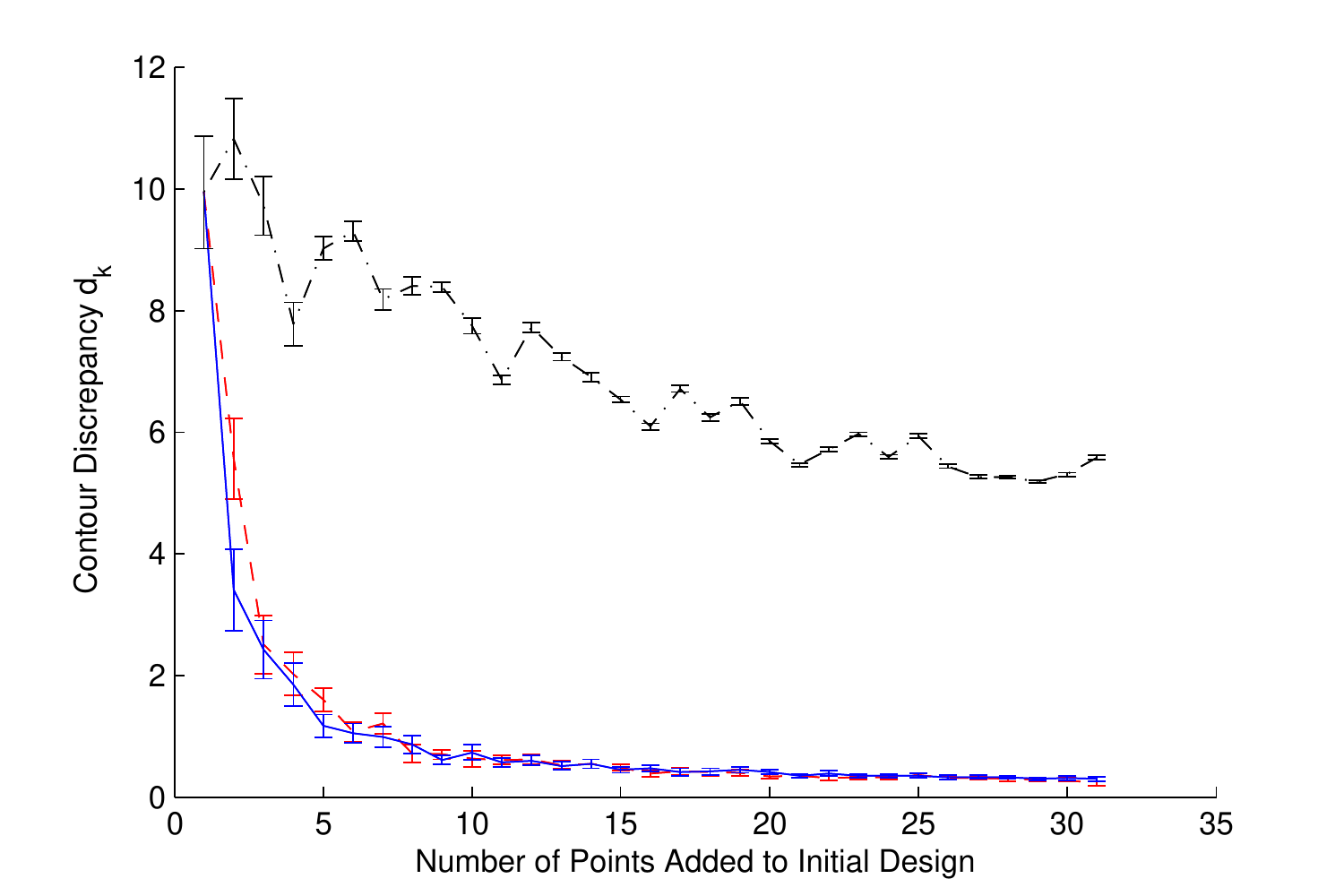}
\caption{Estimated BNB(blue), GA(red), and static(black) contour divergence versus number of new points added to initial design for the Branin test function.}
\label{branincontourdiv}
\end{figure}

A comparison of a static design of size $n_0 + k$ and the sequential design with EI (10) optimized using BNB and GA when estimating a contour of height $y = 45$ is presented in Figure~\ref{branincontourdiv}. Their performance is measured in terms of the divergence between the true and estimated contour, given by
\begin{equation}
d_k=\sqrt{\frac{1}{n} \sum _{i=1}^m (\hat{y}_k(x_{c,i})-a)^2},
\label{contour_divergence}
\end{equation}
where $\{x_{c,i}, i=1,...,m\}$ denotes the discretized true contour at $y=45$, and $\hat{y}_k(\cdot)$ denotes the estimated process value after adding $k$ new points or from a static design of size $n_0+k$. The results are averaged over $100$ random maximin Latin hypercube initial designs of size $n_0=20$ each, and for each realization, we add $n_{new}=30$ additional trials.

Figure \ref{branincontourdiv} displays the divergence $d_k$ for the Branin function with contour at $y=45$. In this simulation BNB and GA methods perform similarly, both significantly better than the static design. The contour discrepancies of the sequential methods approach 0 quickly; an indication that this contour is relatively simple to locate. Since the contour at $y=45$ is in the very bottom left corner of the design region (see Figure~5), the maximin Latin hypercube design is less likely to place points very near the contour $y=45$, which results in poor performance for the static method.\\

\begin{figure}[!h] \centering
\includegraphics[scale=0.8]{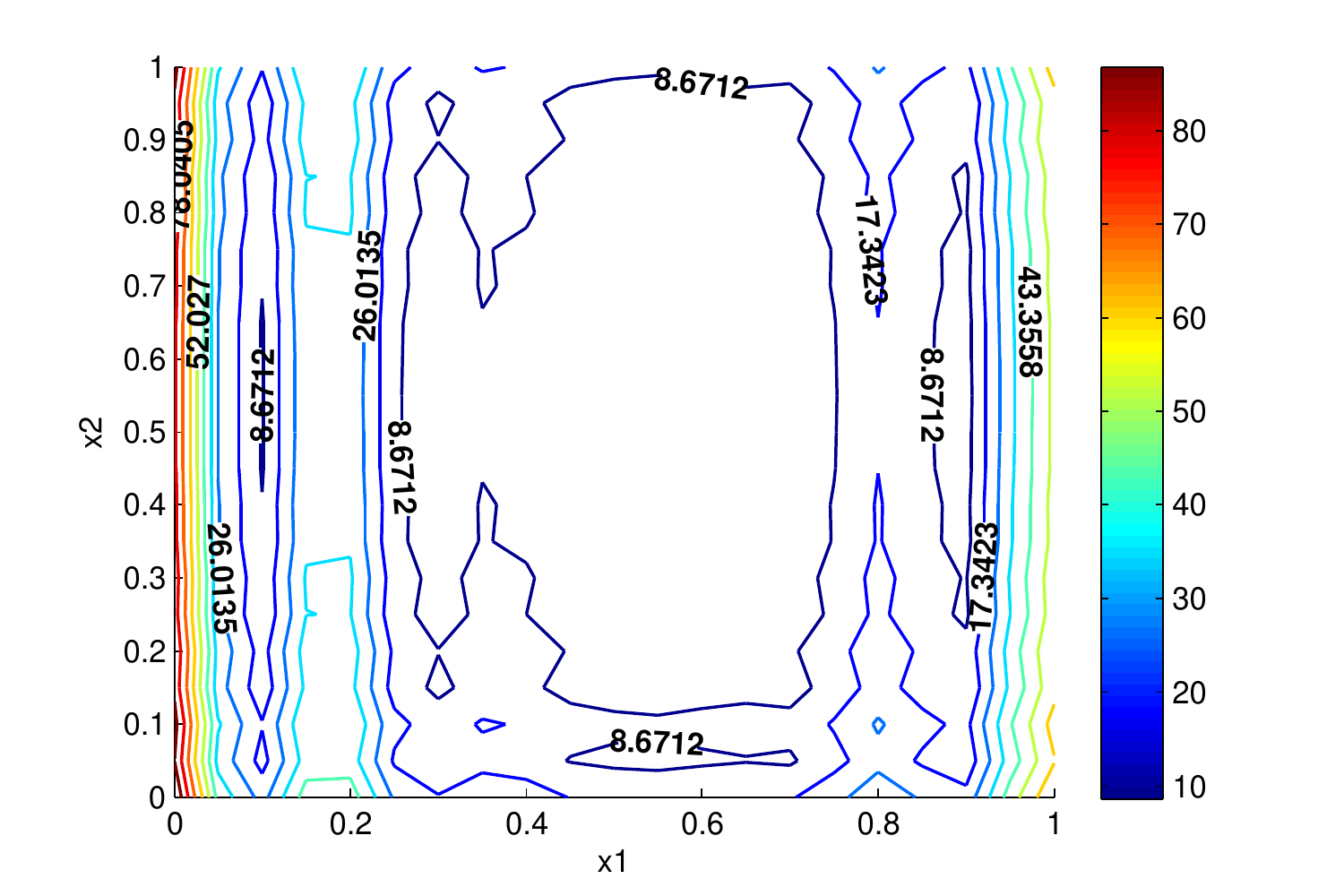}
\caption{2D Levy function contour plot rescaled to $[0,1]^2$}
\label{levytruecontour}
\end{figure}

\textbf{Example~2 - Levy 2D.} Suppose the computer model output $y = f(x_1,x_2)$ is generated using the Levy function given by
\begin{equation*}
f(x_1,...,x_d) = \sin^2(\pi w_1)+ \sum_{k=1}^{d-1}(w_k-1)^2[1+10 \sin^2(\pi w_k+1)]+(w_d-a)^2,
\label{levyfunction}
\end{equation*}
where $w_k=1+(x_k-1)/4$ and $x_k \in [-10,10]$ for $k=1,...,d$. We rescale the inputs $x = (x_1,...,x_d)$ to $[0,1]^d$. For the  two-dimensional Levy function (see Figure \ref{levytruecontour}), the global maximum, $y_{max} =95.4$, and minimum, $y_{min} = 0$, are attained at $(x_1,x_2)=(0,0)$ and $(x_1,x_2)=(0.55,0.55)$.

\begin{figure}
  \centering
  \subfloat[Minimum]{\label{levymaxminMin}\includegraphics[height=3in,width=2.5in]{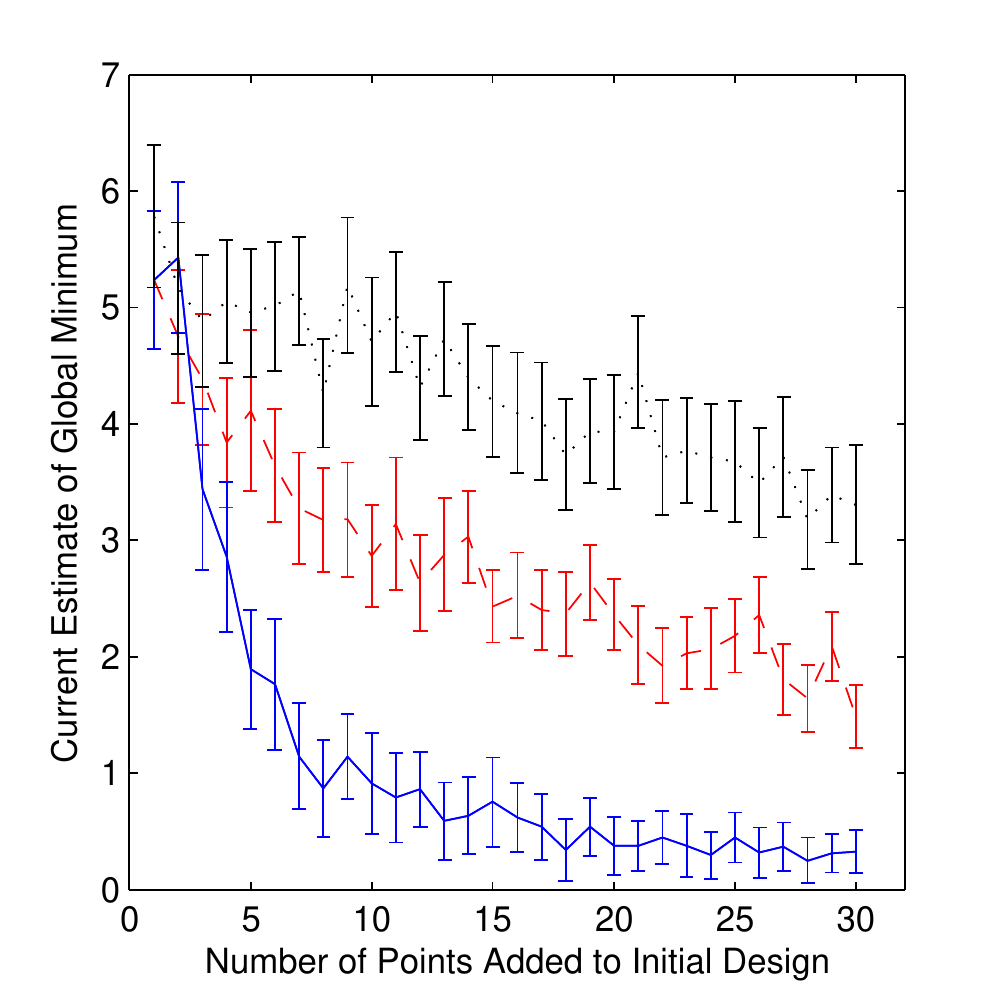}}
  \subfloat[Maximum]{\label{levymaxminMax}\includegraphics[height=3in,width=2.5in]{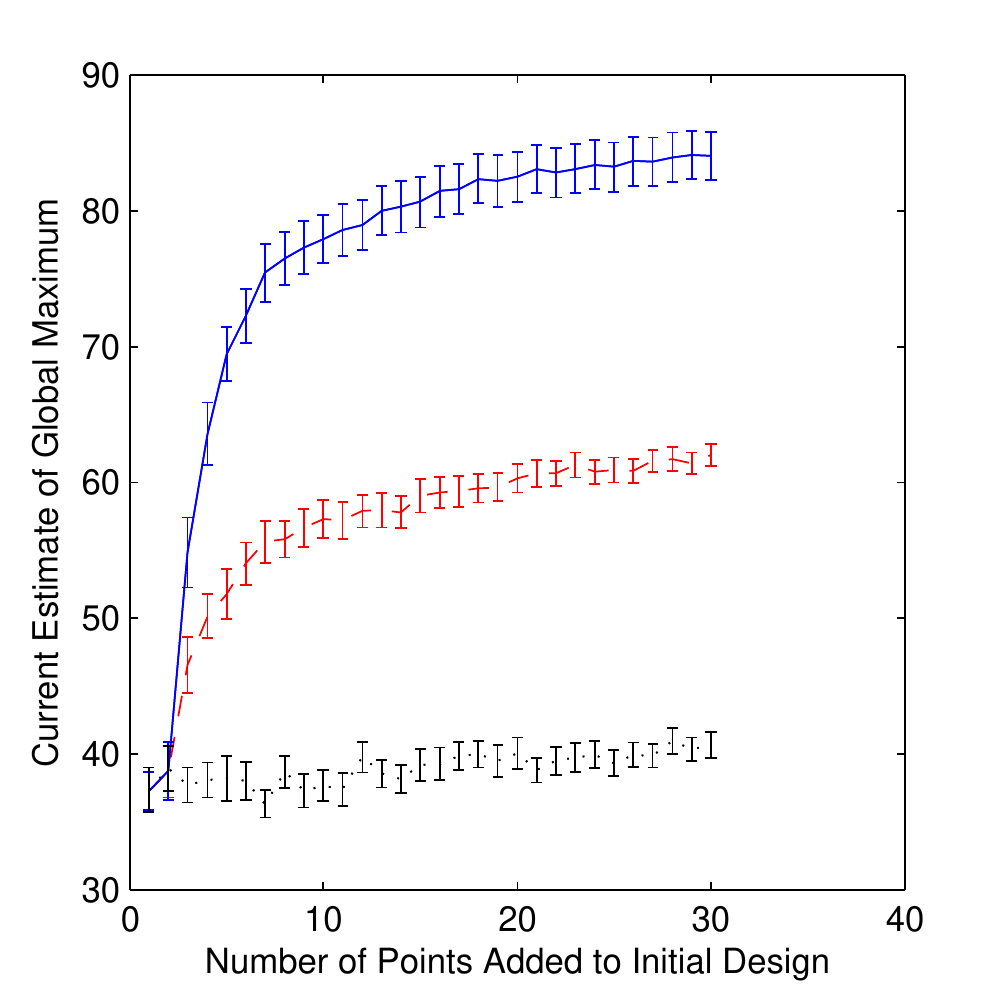}}
  \caption{Estimated optima for Levy 2D function (true minimum = 0, maximum = 95.4); BNB (blue - solid), GA (red - dashed), static (black - dash-dotted).}
  \label{levymaxmin_plot}
\end{figure}

\begin{figure} \centering
\includegraphics[scale=0.6]{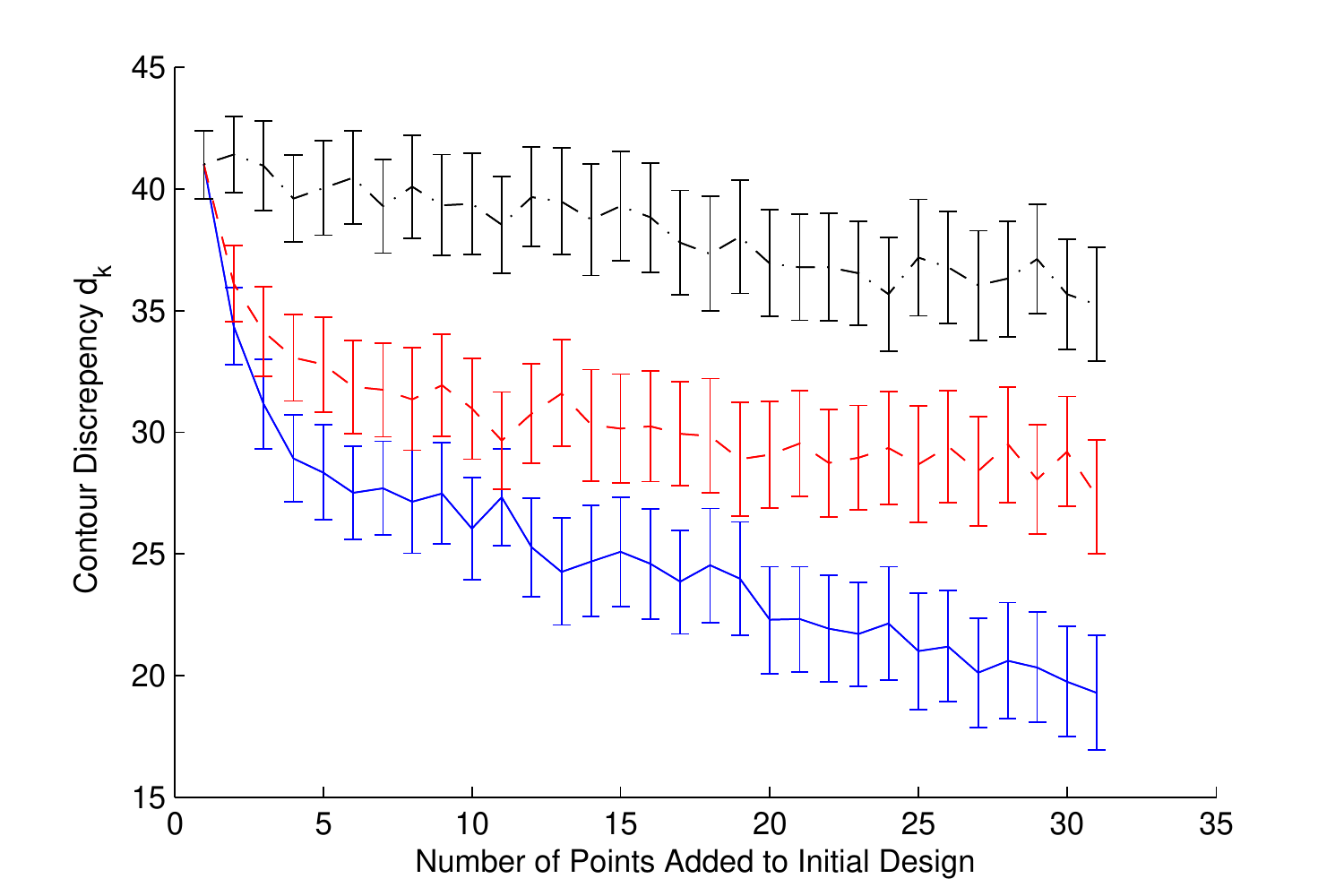}
\caption{Estimated contour divergence comparison for the 2D Levy function ($y=70$); BNB (blue - solid), GA (red - dashed) and static (black - dash-dotted).}
\label{levycontourdivergence}
\end{figure}

Figure \ref{levymaxmin_plot} presents the long run comparison of the BNB and GA optimizers for maximizing the EI criterion (\ref{ei_maxmin}) developed for simultaneous estimation of the global maximum and minimum. As in Example~1, the results are averaged over $100$ realizations with initial designs of size $n_0=20$ and $n_{new}=30$ additional trials. For this example, BNB is a clear winner and leads to better estimates of the global optima in much fewer simulator runs for both the maximum and minimum.  As expected, GA performs significantly better than the static design.
Figure~\ref{levycontourdivergence} presents the divergence comparison of the sequential design with the two optimizers as well as a static design when estimating the contour at $y=70$. Here also, each initial design was of size $n_0=20$, $n_{new}=30$ new trials were added sequentially, and the results are averaged over $100$ initial random maximin Latin hypercube designs.  As in the maximum and minimum case, the sequence of points added by the BNB (blue) optimization of EI (10) lead to quicker convergence of the contour divergence compared to GA (red).  On average the static (black) designs provide relatively minimal improvement to the contour estimate.

\textbf{Example 3 - Levy 4D.} We now use an example of a simulator with four-dimensional inputs (Levy function) to demonstrate that careful optimization of the EI criterion can save significant amount of simulator runs. The simulation results use 30 initial design points chosen from random maximin Latin hypercube, and 20 additional points added sequentially. Comparisons using this higher dimensional function may highlight the differences between the two methods of optimization. BNB and GA were each given a budget of 3000 evaluations of the EI criterion.  The true global maximum is approximately 255 at $(0,0,0,0)$, and the minimum is 0 at $(0.55,0.55,0.55,0.55)$.

\begin{figure}[!h]  \centering
  \subfloat[Estimated Minimum]{\label{levy4DmaxminMin}\includegraphics[height=2.85in,width=2.5in]{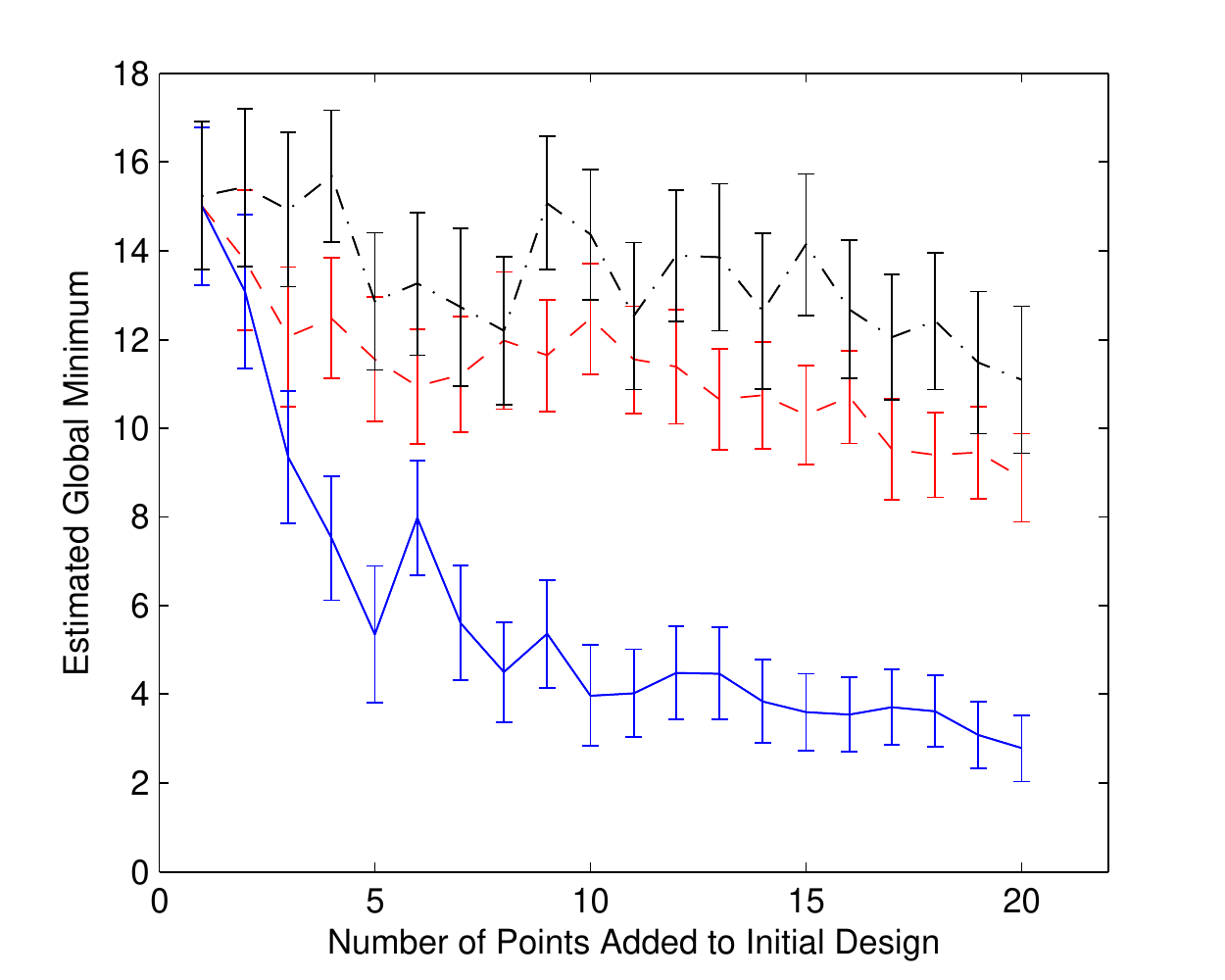}}
  \subfloat[Estimated Maximum]{\label{levy4DmaxminMax}\includegraphics[height=2.85in,width=2.5in]{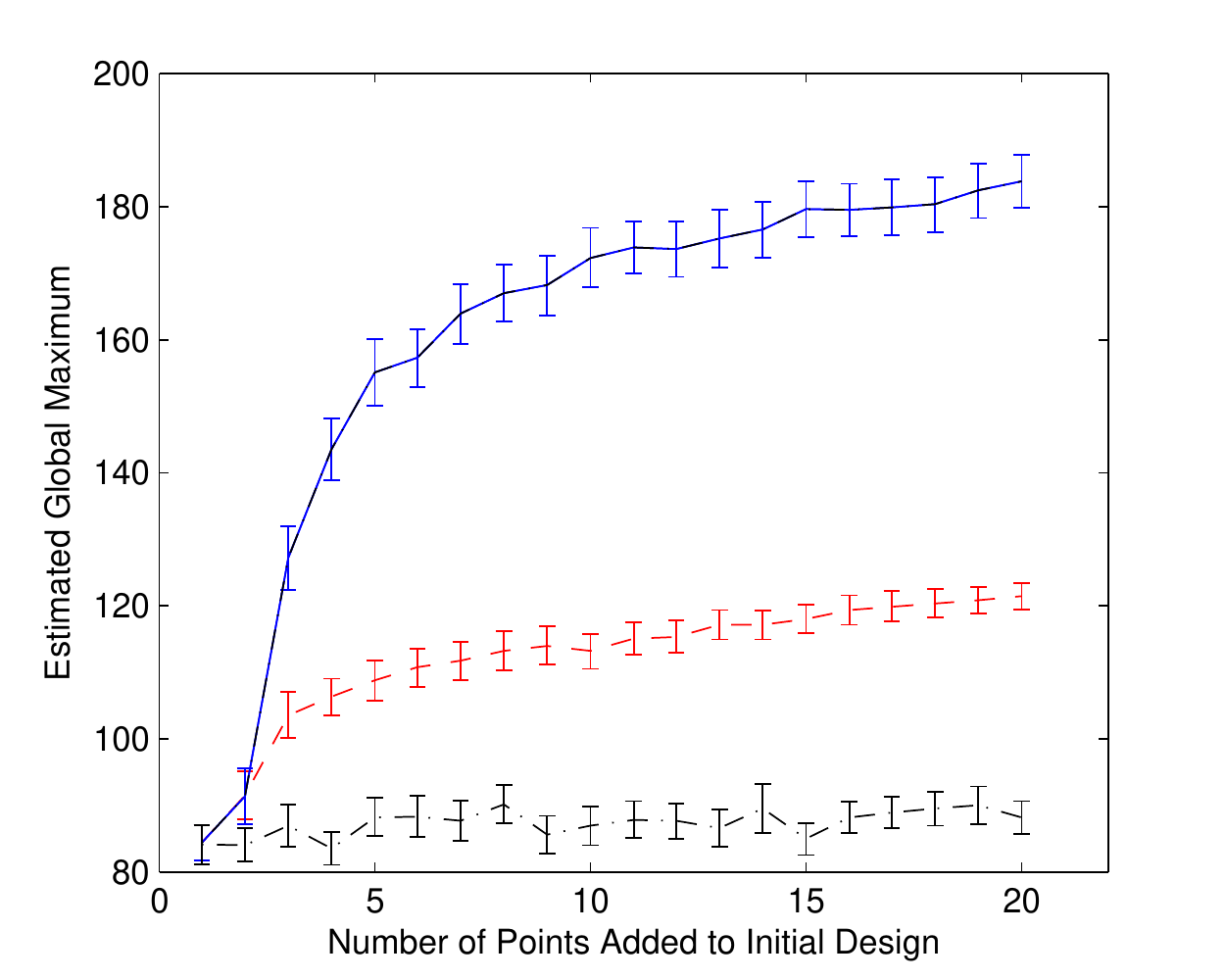}}
  \caption{Estimated optima for Levy 4D function (true minimum = 0, maximum = 255); BNB (blue - solid), GA (red - dashed) and static (black - dash-dotted).}\label{levy4dmaxmin}
\end{figure}

Figure \ref{levy4dmaxmin} displays the comparison of our BNB algorithm with the GA and static designs in simultaneously estimating the maximum and minimum of the 4D Levy function. The results are averaged over 100 random maximin Latin hypercube initial designs of size $n_{0}=30$. BNB begins to locate both the minimum and maximum much more quickly than GA, which is in turn better than the static design. BNB is the only method that gets close to the maximum in 20 additional trials. All three methods are slow to locate the global maximum, a reflection of the complexity of the Levy function.

\begin{figure}[!h] \centering
\includegraphics[width=0.8\textwidth]{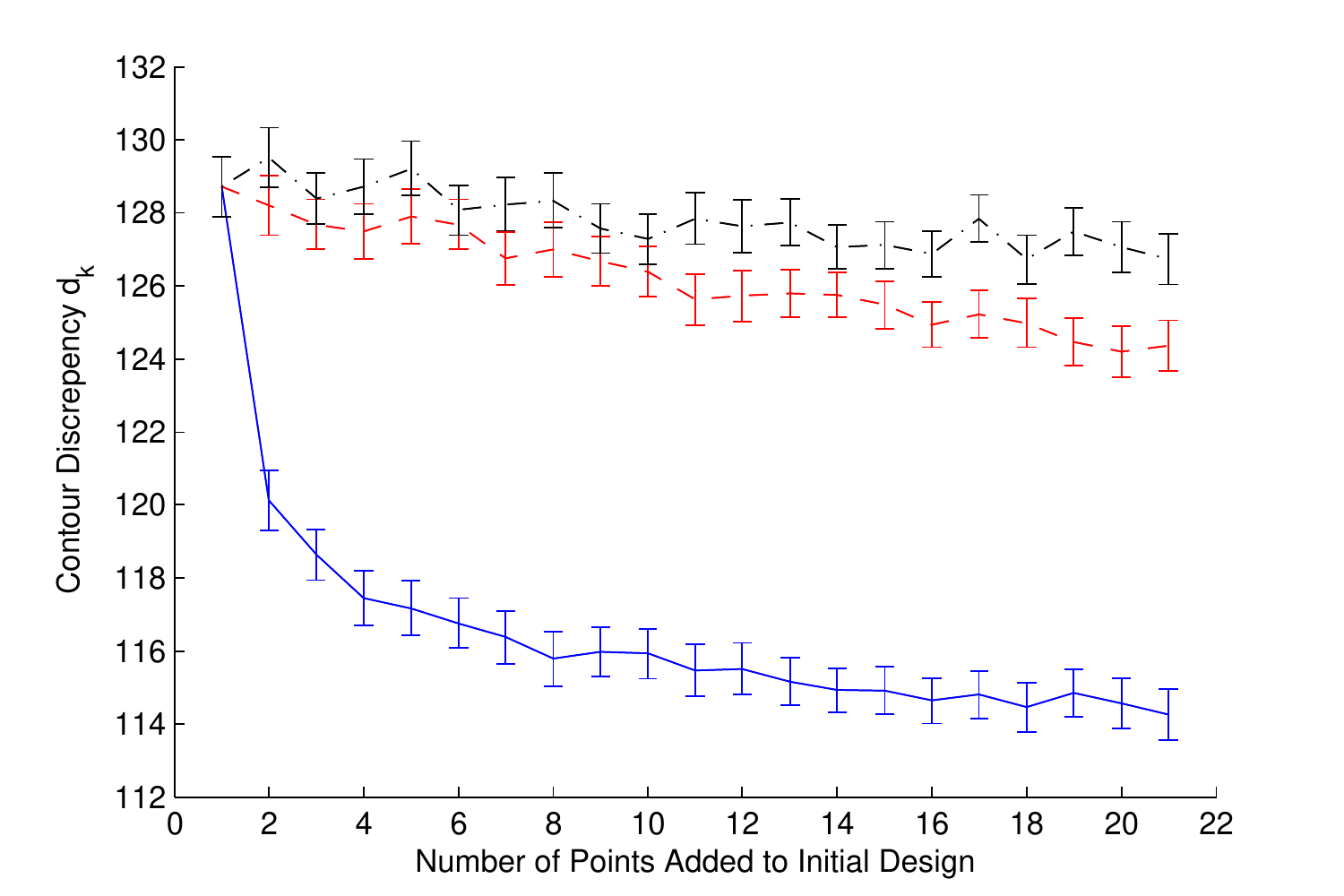}
\caption{Estimated contour divergence comparison for the 4D Levy function ($y=180$); BNB (blue - solid), GA (red - dashed) and static (black - dash-dotted).} \label{levy4dcontour}
\end{figure}

As before, we simulate the estimation of the $y=180$ contour of the 4D Levy function for BNB, GA, and static designs.  The number of initial points $n_{0}$ was 30 and 20 new points were added. The results are averaged over 100 random maximin Latin hypercube designs. Figure \ref{levy4dcontour} displays the contour divergence $d_k$ for the three designs - sequential with EI (10) optimized using BNB, GA and the static design. The new points selected by the BNB  algorithm allow estimation of the 4D Levy $y=180$ contour more accurately than the other two methods, decreasing the contour divergence much more rapidly after only one point was added.

As illustrated in these examples, the branch and bound
algorithm is quite competitive. For 2D and 4D Levy functions, BNB is superior to GA and static in estimation of contours as well as simultaneous estimation of the maximum and minimum.  In estimating the maximum and minimum of the Branin function BNB locates the maximum much more quickly than the other methods, and all 3 methods are similar in estimating the minimum.  For contours of the Branin function GA performs on par with BNB, both of which are preferable to a static design.  The Branin function is relatively smooth, and hence optimization of EI criteria is relatively simple, which may explain the absence of differences between these methods in some cases.

\section{Discussion} \label{conclusion}

We have generalized the expected improvement criterion for finding the maximum and minimum (simultaneously) as the features of interest, and demonstrated the use of this infill criterion on the Levy and Branin test functions.
Also we have developed branch and bound algorithms for contour and maximum/minimum estimation, compared them with genetic algorithms and static designs, and have shown that the BNB outperforms the other two optimization approaches. This study demonstrate that careful optimization of the EI criterion can save significant amount of simulator runs.

Note that we have used a `stochastic version' of the BNB algorithm, and the approximation comes in from estimating the bounds of $\hat{y}(\mathbf{x})$, $\hat{s}(\mathbf{x})$ based on points in each rectangle $\mathcal{Q} \in \mathcal{Q}_{init} = [0,1]^d$. Further work should be done to investigate maximization of the proposed EI criteria using a non-stochastic branch and bound algorithm.



\bibliographystyle{agsm}
\bibliography{Franey_Ranjan_Chipman}

\end{document}